\documentclass[a4paper,11pt]{article}
\usepackage{pos}

\title{Studies of a muon-based mass sensitive parameter for the IceTop surface array}
\ShortTitle{Mass sensitive parameter for IceTop}

\author{The IceCube Collaboration \\{\normalsize \normalfont(a complete list of authors can be found at the end of the proceedings)}}

\emailAdd{donghwa.kang@kit.edu}
\emailAdd{sally-ann.browne@student.kit.edu}
\emailAdd{andreas.haungs@kit.edu}

\abstract{IceTop is the surface instrumentation of the IceCube Neutrino Observatory at the South Pole. It is designed to measure extensive air showers of cosmic rays in the primary energy range from PeV to EeV. Air showers induced by heavier primary particles develop earlier in the atmosphere and produce more muons observable at ground level than lighter cosmic rays with the same primary energy. Therefore, the fraction of muons to all charged particles measured by IceTop characterizes the mass of primary particles. This analysis seeks a muon-based mass sensitive parameter by using the charge signal distribution for each individual cosmic ray event. In this contribution we present the analysis method for the mass-sensitive parameter and our studies of its possible application to the measurement of cosmic ray mass composition with the IceTop surface array.

\vspace{4mm}
{\bfseries Corresponding authors:}
Donghwa Kang$^{1*}$, Sally-Ann Browne$^{2}$, Andreas Haungs$^{1}$\\
{$^{1}$ \itshape Karlsruhe Institute of Technology, Institute for Astroparticle Physics, Karlsruhe, Germany\\
$^{2}$ \itshape Karlsruhe Institute of Technology, Institute of Experimental Particle Physics, Karlsruhe, Germany}\\[4mm]
$^*$ Presenter

\FullConference{37$^{\rm{th}}$ International Cosmic Ray Conference (ICRC 2021)\\
		July 12th -- 23rd, 2021\\
		Online -- Berlin, Germany}

}

\begin{document}
\maketitle

\section{Introduction}
Investigations on the mass composition of cosmic rays in the primary energy range of PeV to EeV are essential to understand their sources and propagation mechanism. This energy range is particularly important, 
because it is believed to be the Galactic to extragalactic transition region.
Air showers induced by heavier primary particles develop earlier in the atmosphere due to their larger cross section for interacting with air nuclei, and produce more muons of lower energies observable at ground level than lighter cosmic rays with the same primary energy.
The fraction of muons to all-charged particles at the observation level characterizes the mass of primary particles, i.e. electron-rich showers (electron stands for all particles of the electromagnetic component) are generated by light primary nuclei and electron-poor showers by heavy nuclei, respectively. 
This work investigates a muon-based mass sensitive parameter for each individual event and its possible application for composition studies of primary cosmic rays.

IceTop \cite{IceTop} is the 1-km$^{2}$ surface array of the IceCube Neutrino Observatory at the geographic South Pole.
It consists of 81 stations with an average spacing of 125 m between neighboring stations above the IceCube strings.  
Each station has a pair of ice-Cherenkov tanks and each tank is instrumented with two Digital Optical Modules (DOMs). 
The measurement of the primary cosmic-ray energy spectrum and composition using three years of coincidence data from IceTop and IceCube was recently published \cite{3yrSpectrum}. The results are consistent with measurements by other experiments yet with significant systematic uncertainties mainly due to the snow overburden above the IceTop tanks and the dependency on composition and hadronic interaction models used for energy assignments.

In general it is reasonable to assume that the muon signal becomes significant in tanks at a large distance from the shower axis. Signals in tanks closer to the shower core are dominated by the electromagnetic components. Considering the charge signal distribution, the so-called muon enriched parameter per individual shower is introduced and estimated, where the original idea comes from Ref. \cite{IceTopMuon}.
This muon-based mass sensitive parameter would make it possible to divide data sets into heavy and light primary groups and therefore the composition studies eventually can be performed.
In addition, using this approach one would be able to reduce systematic uncertainties, in particular, those depending on composition assumption, since the energy spectrum of cosmic rays can be reconstructed using the ratio between the shower size measured at 125 m from the shower axis (S125) and the estimated muon parameter.

In this contribution we describe a detailed analysis technique for estimating the mass-sensitive parameter and its possible application to the measurement of cosmic ray mass composition with the IceTop surface array.

\section{Data sets and quality cuts}
The analysis discussed here uses the standard IceCube Monte-Carlo simulations based on the hadronic interaction model of Sibyll 2.1 \cite{Sibyll21}. Four different primary particles (proton, helium, oxygen, iron) are simulated with 20,000 showers each with a spectral index of -1 from the energy range of $10^{14}$ to $10^{17}$ eV. The simulations are weighted to the spectral index of -2.7. The showers generated by CORSIKA \cite{corsika} are propagated into the detector, in which the individual tank response is simulated by Geant4 \cite{geant4} including snow overburden and barometric conditions. About 10\% of 2012 data measured by IceTop was used as a test sample for the comparison with simulations.

All standard quality cuts are applied \cite{3yrSpectrum}. The shower reconstruction algorithm should converge and the number of stations must be larger than 5. The core position of the air shower must be within the radius of 400 m. The zenith angle range from 0 to 18$^{\circ}$
is applied for the present analysis. Cuts on the shower size log$_{10}$(S125/VEM) from 1 to 1.5 are applied, which correspond to the true energy from 10$^{16}$ to 10$^{16.5}$ eV. 

\section{Analysis}
In IceTop the shower size (S125), which is a signal at a reference distance of 125 m from the shower axis, is estimated with high accuracy of $\Delta$log$_{10}$(S125/VEM) $\approx$ 0.05 \cite{IT73}. However, there is no muon number estimated for individual shower. Thus, the first step to perform this analysis is to estimate the muon enriched parameter. Note that this parameter contains still a small contamination of electromagnetic components due to the muon selection efficiency. Thus, we call it the muon enriched parameter.
\begin{figure}[t!]
  \begin{center}
    \includegraphics[width=0.45\textwidth]{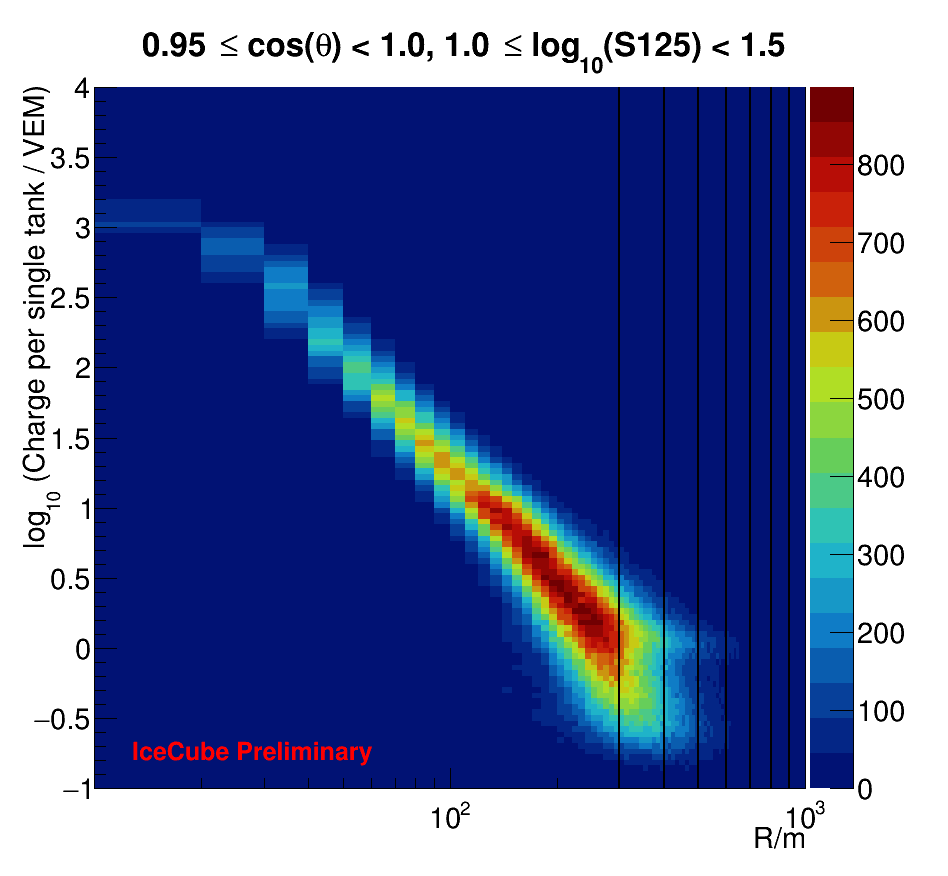}
    \includegraphics[width=0.54\textwidth]{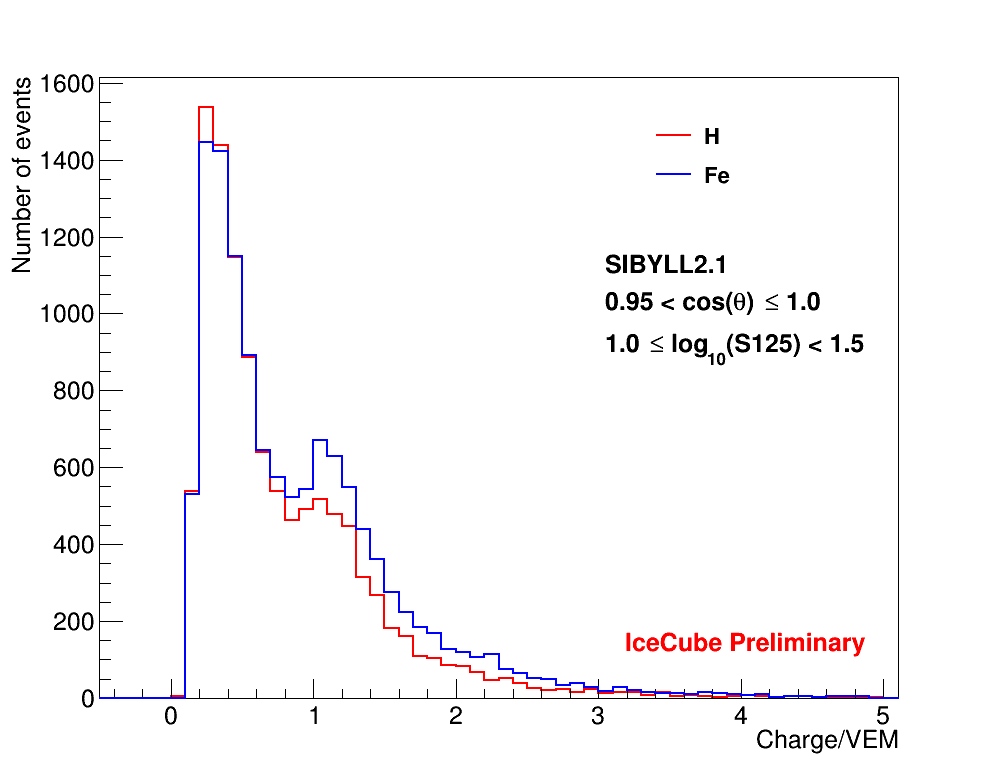}
    \caption{Left: The 2-dimensional histogram of charge signal per single tank as a function of distance from the shower axis. The charge signal is in the unit of Vertical Equivalent Muon (VEM). Both HLC and SLC signals are used. The z-axis is the number of events.
    Right: The charge distribution for a fixed distance from 350 to 450 m for the proton and iron primaries. The first distinct peak stems from electromagnetic components and the second one is caused by muons at 1 VEM.}
\label{fig1}
\end{center}
\end{figure}

\subsection{Charge distribution}
The 2-dimensional charge distribution for 10\% of 2012 IceTop data as a function of distance can be seen on the left in Fig. 1. The charge is in the unit of Vertical Equivalent Muon (VEM), which is a signal initiated by a muon passing vertically through the tank.
For this analysis both Hard Local Coincidence (HLC) and Soft Local Coincidence (SLC) signals are used.
The HLC signals occur when two tanks from the same station are triggered within a time window of 1 $\mu$s,
whereas the SLC signals do not have such triggered neighbor.
The charge signal decreases with increasing lateral distances and there is a peak at around 1 VEM, which is mainly created by muons. This muon signal mainly comes from SLC signals, since a SLC signal is likely to occur at large distance due to its small trigger probability. To illustrate the shape of the charge distribution at a fixed distance more detailed, it is divided by a distance bin of 100 m.

Figure 1 (right) is the vertical slice of the 2-dimensional charge distribution, e.g. for a distance from 350 to 450 m. There are two distinct peaks: the first peak stems from electromagnetic components of showers and the second one is the muon peak at 1 VEM. The muon peak is more dominant for the case of iron showers. The reason is that for the same energy the iron-induced air showers produce more muons observable at the ground level.
In order to select muon signals for the muon parameter, a straightforward cut is applied to the charge distribution. The value of 0.7 VEM for the charge cut is obtained by the valley position from the signal calibration of a single tank \cite{ICRC2011}.
This cut implies that the charges above the cut still contain some contamination of electromagnetic background. In general, to suppress the background one has to fit both the signal and the background distribution and subtract the background. However, this fitting procedure does not work properly for each individual shower due to low statistics. Therefore, we use this cut and the background will be corrected by the muon correction factor later (section 3.3).
The charge distributions from data and Monte Carlo are in agreement with each other, however, a detailed comparison of data and Monte Carlo with the composition assumption, e.g. the H4a model, is under investigation \cite{SallyThesis}.

\begin{figure}[b!]
  \begin{center}
    \includegraphics[width=0.50\textwidth]{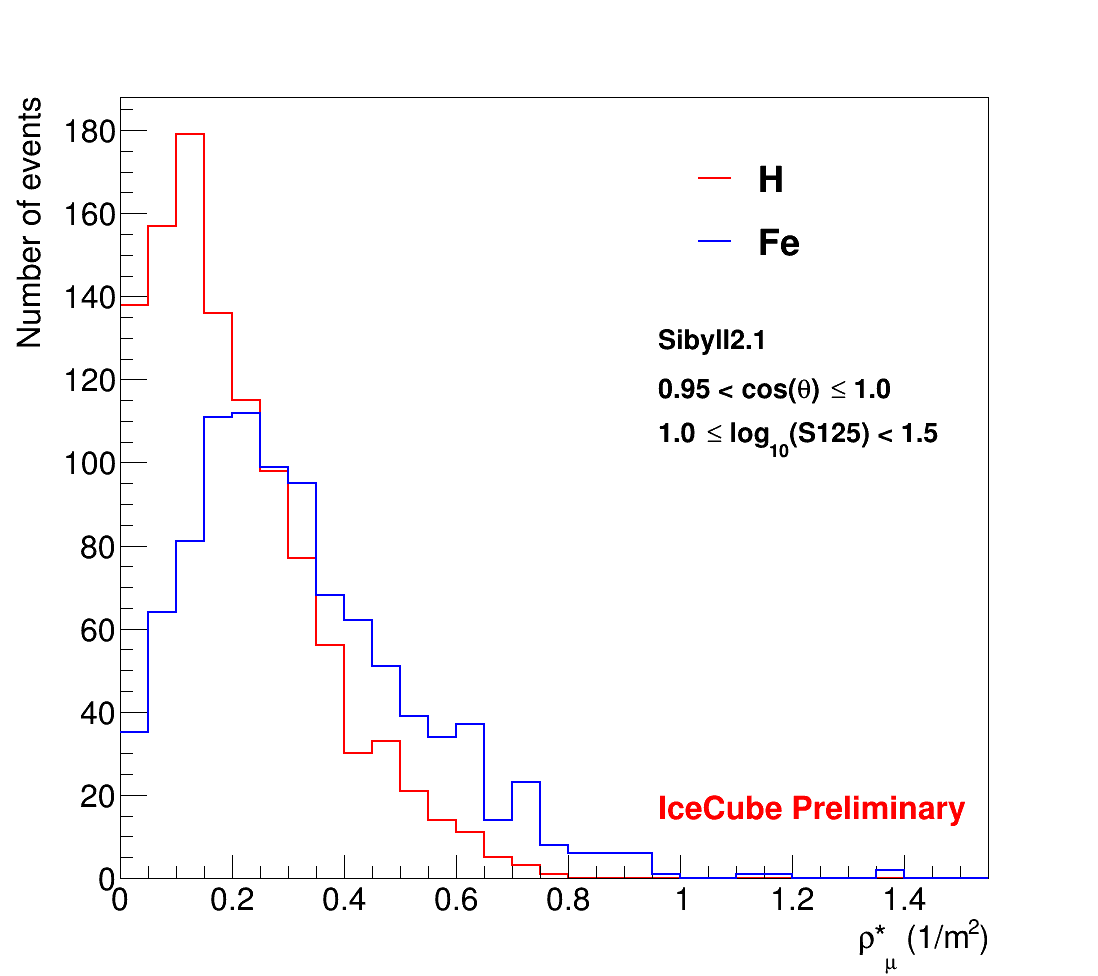}
    \caption{Estimated muon-based parameter ($\rho^{*}_{\mu}$) at a fixed distance from 350 to 450 m, event by event, for proton (red) and iron (blue) primaries.}
\label{fig2}
\end{center}
\end{figure}

\subsection{Estimation of muon parameter}
Considering the charge signal distribution, the parameter sensitive to the muon content was estimated, which is the sum of the charge signals ($\geq$ 0.7 VEM) divided by the total number of tanks and the area of the tanks at a fixed distance from the shower axis:
\begin{equation}
  \rho^{*}_{\mu} = \frac{\sum{charge}}{N_{tank} \times A_{eff}},
\end{equation}
where the effective area of a tank $A_{eff}$ is calculated with the radius of 91 cm.
Figure 2 shows the estimated muon-based parameter ($\rho^{*}_{\mu}$) per individual showers, i.e. event by event, at a reference distance of 400 m, assuming proton and iron primaries. The distribution of the muon parameter for the iron shower is slightly shifted to the right, which implies that the value of the muon parameter at 400 m is larger for iron. $\rho^{*}_{\mu}$ is contaminated by electromagnetic backgrounds, so that the correction has to be done by using the true muon density.

\subsection{Muon correction factor}
In order to derive the muon correction factor, true muon signals were selected by using the true muon information. The charge pulses, which were generated in the tank after propagating CORSIKA output particles, are used. However, this charge is raw data, i.e. it is not calibrated and the snow is not corrected. Then the fraction of the signals caused by muons in each tank is calculated. If the signal has the muon fraction larger than half of the total charge signal ($f_{\mu} > 0.5$), then this is considered as true muon signal. This cut for the true muon selection is actually obtained by the optimization of the muon purity and the statistical point of view.

\begin{figure}[t!]
  \begin{center}
    \includegraphics[width=0.49\textwidth]{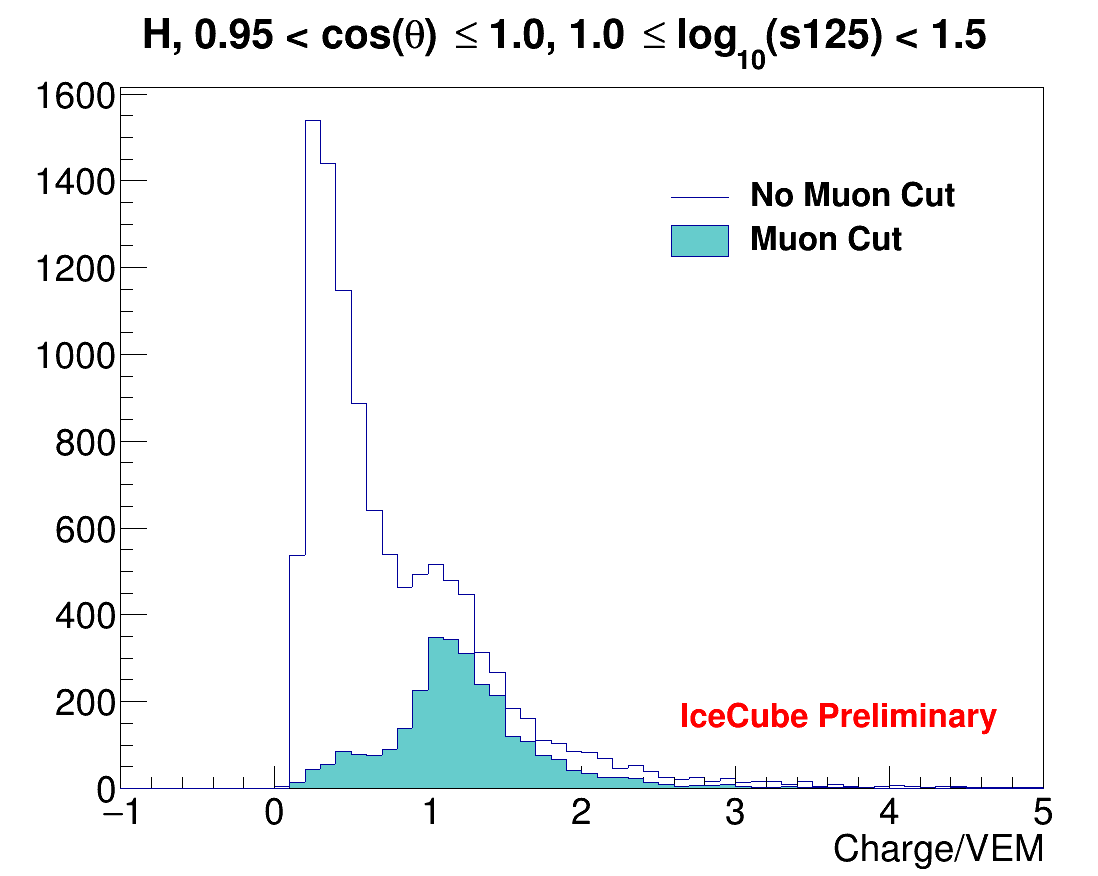}
    \includegraphics[width=0.49\textwidth]{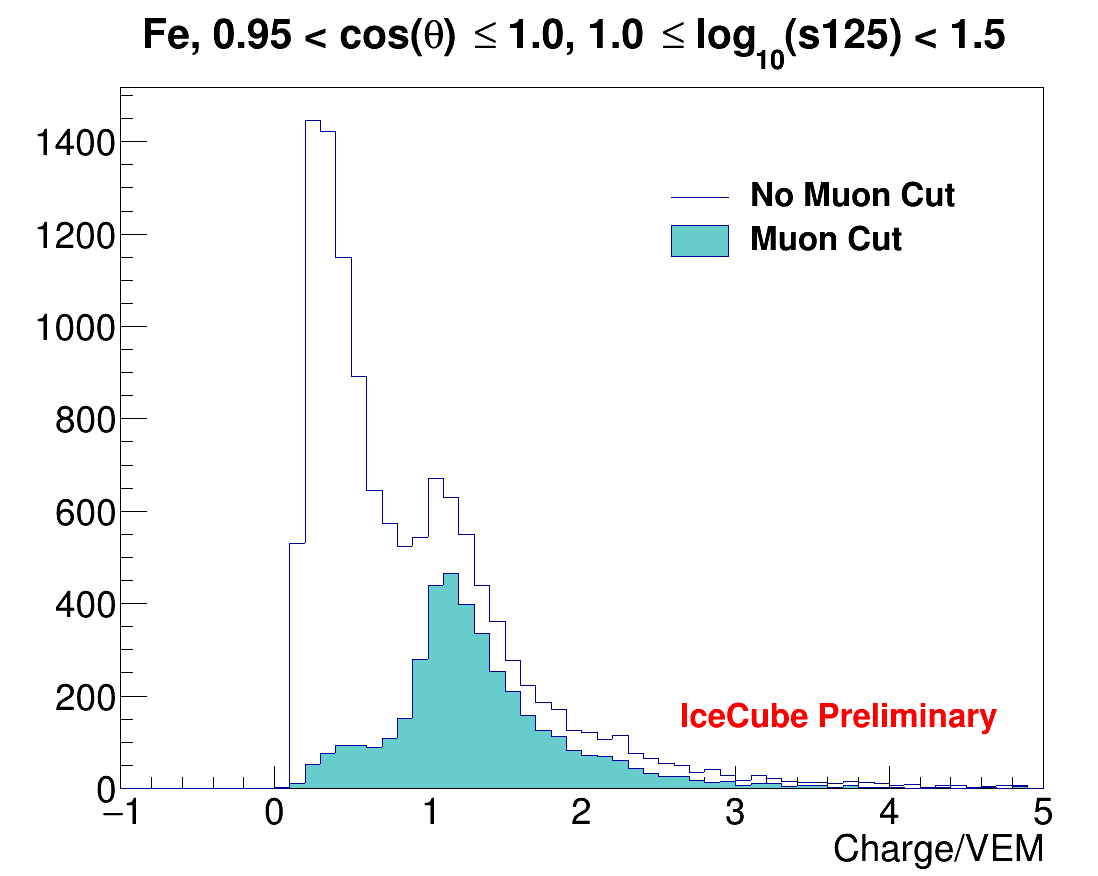}
    \caption{True muon contributions to used charge for the muon estimation for proton (left) and iron (right) primaries. The blue filled histograms are the charge distributions after the true muon selection.}
\label{fig3}
\end{center}
\end{figure}
Figure 3 presents true muon contributions to 
the muon estimation for proton (left) and iron (right) primaries. After the muon selection, a large amount of electromagnetic components are reduced and the muon peak appears significantly. About 30\% of charge signals remained for proton and iron primaries, respectively. By using the same procedure, the parameter for the true muon signal was estimated.
\begin{figure}[t!]
  \begin{center}
    \includegraphics[width=0.45\textwidth]{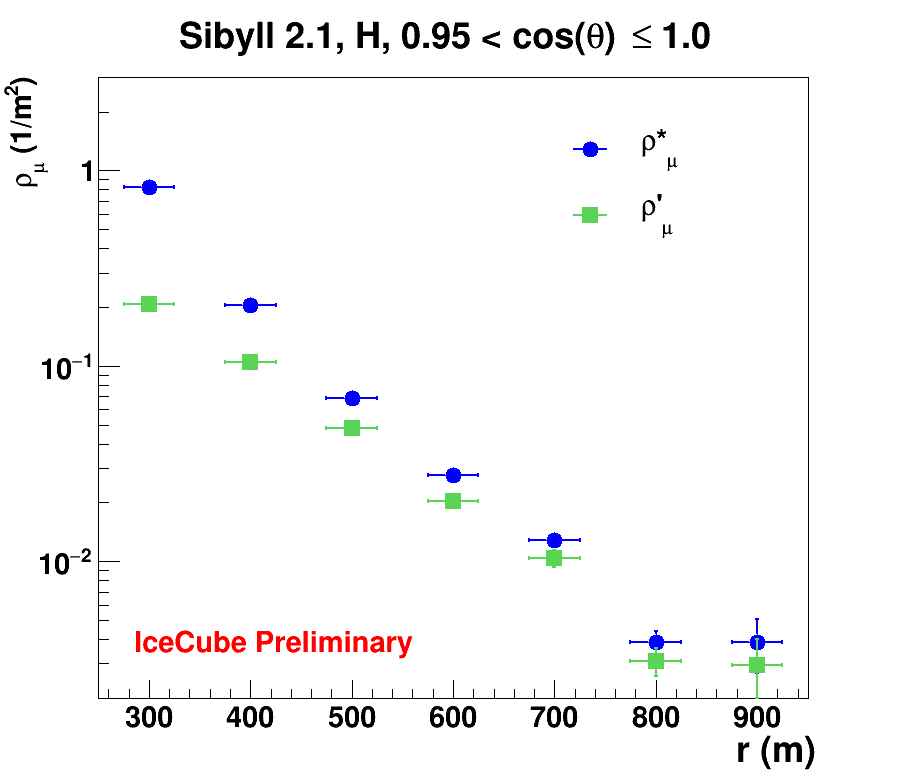}
    \includegraphics[width=0.45\textwidth]{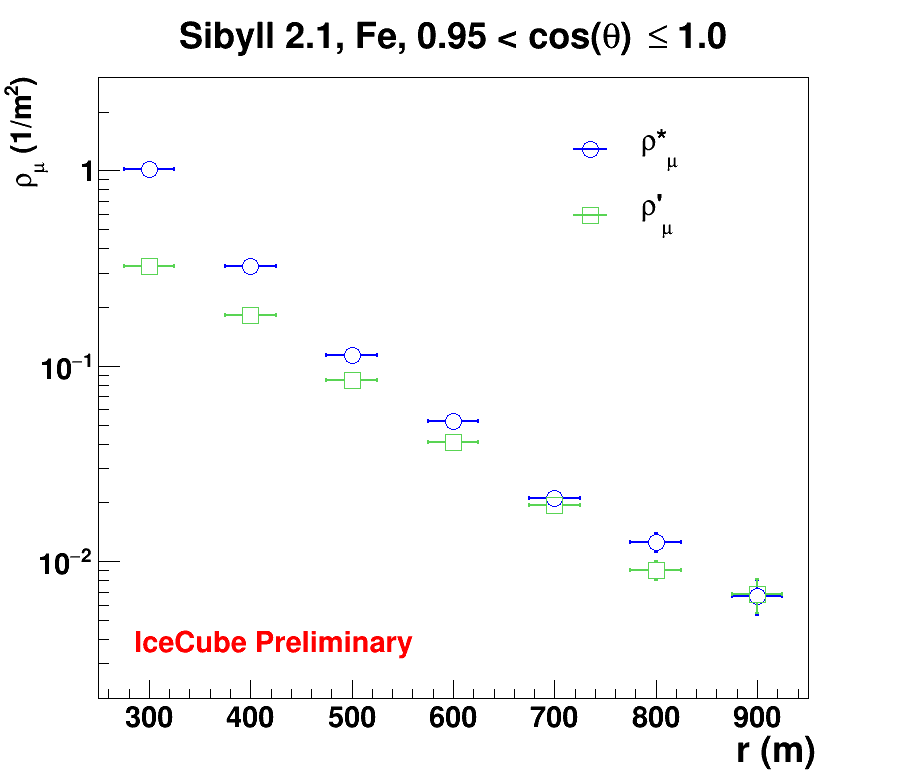}
    \caption{The muon parameter before and after the muon selection as a function of distances for proton (left) and iron (right) assumptions. $\rho^{*}_{\mu}$(blue) and $\rho^{'}_{\mu}$(green) are the muon parameters before and after the muon selection, respectively.}
\label{fig4}
\end{center}
\end{figure}
The estimated muon parameter before and after muon selection can be seen in Fig. 4. The difference becomes smaller at larger distances. It means that the muons are more dominant than the electromagnetic components at larger distances. So, this difference was taken into account as the correction factor.
$\rho^{*}_{\mu}$(blue) and $\rho^{'}_{\mu}$(green) are the muon parameters before and after the muon selection, respectively.
\begin{figure}[b!]
  \begin{center}
    \includegraphics[width=0.5\textwidth]{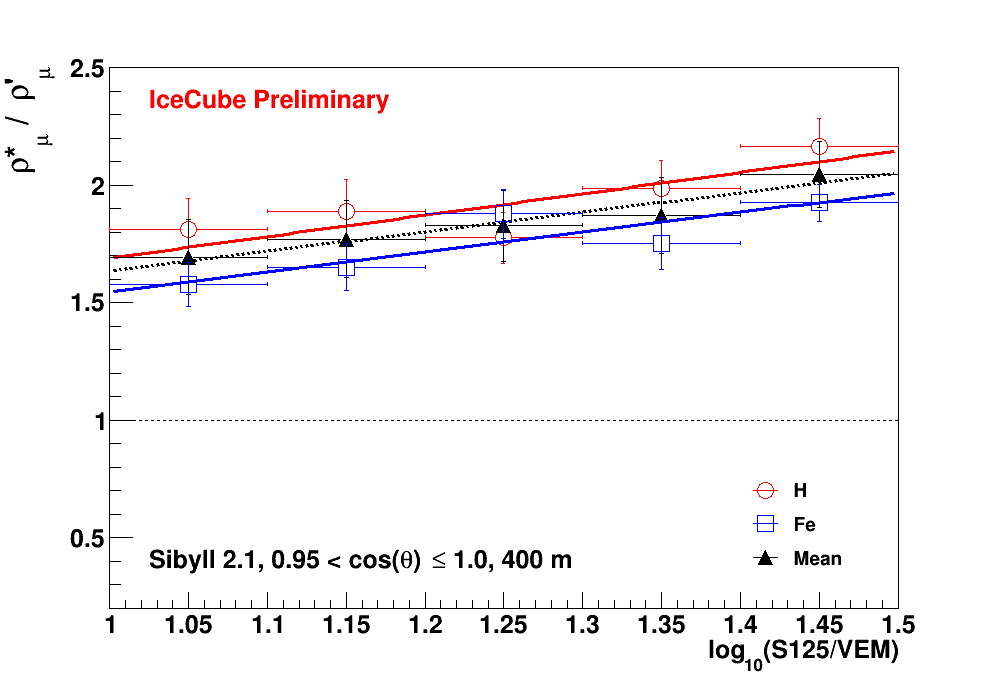}
    \caption{Energy-dependent muon correction factor (CF) for proton and iron primaries, showing the lines of best fit.
$\rho^{*}_{\mu}$ is corrected by CF per individual showers to obtain the $\mu$-based parameter. The average value (black triangle) from proton and iron distributions is applied to experimental data.
}
\label{fig5}
\end{center}
\end{figure}
The muon Correction Factor (CF) is defined by the ratio between the reconstructed $\rho^{*}_{\mu}$ and the true $\rho^{'}_{\mu}$ muon parameter as shown in Fig. 5.
Red circles are for the assumption of proton induced showers and blue squares for iron induced showers.
This correction factor becomes obviously smaller for larger lateral distances, e.g. 600 m.
Since this correction factor depends on the energy, i.e. the energy estimator S125, the fitting with a linear function $f(x) = p0_{H,Fe} \cdot x + p1_{H,Fe}$ was performed.
Coefficients of the linear fitting are $p0_{H} = 0.78 \pm 0.54$ and $p1_{H} = 0.91 \pm 0.43$ for proton-induced air showers, and $p0_{Fe} = 0.71 \pm 0.40$ and $p1_{Fe} = 0.84 \pm 0.32$ for iron-induced showers. For experimental data an average value from proton and iron primaries in each S125 bin is used as a correction factor: $p0_{data} = 0.81 \pm 0.63$ and $p1_{data} = 0.83 \pm 0.49$.
Along with this, the correction function depending on the energy was obtained from the fit results. So, the estimated muon parameter was multiplied by the inverse of the correction factor event by event. 
\begin{figure}[t!]
  \begin{center}
    \includegraphics[width=0.49\textwidth]{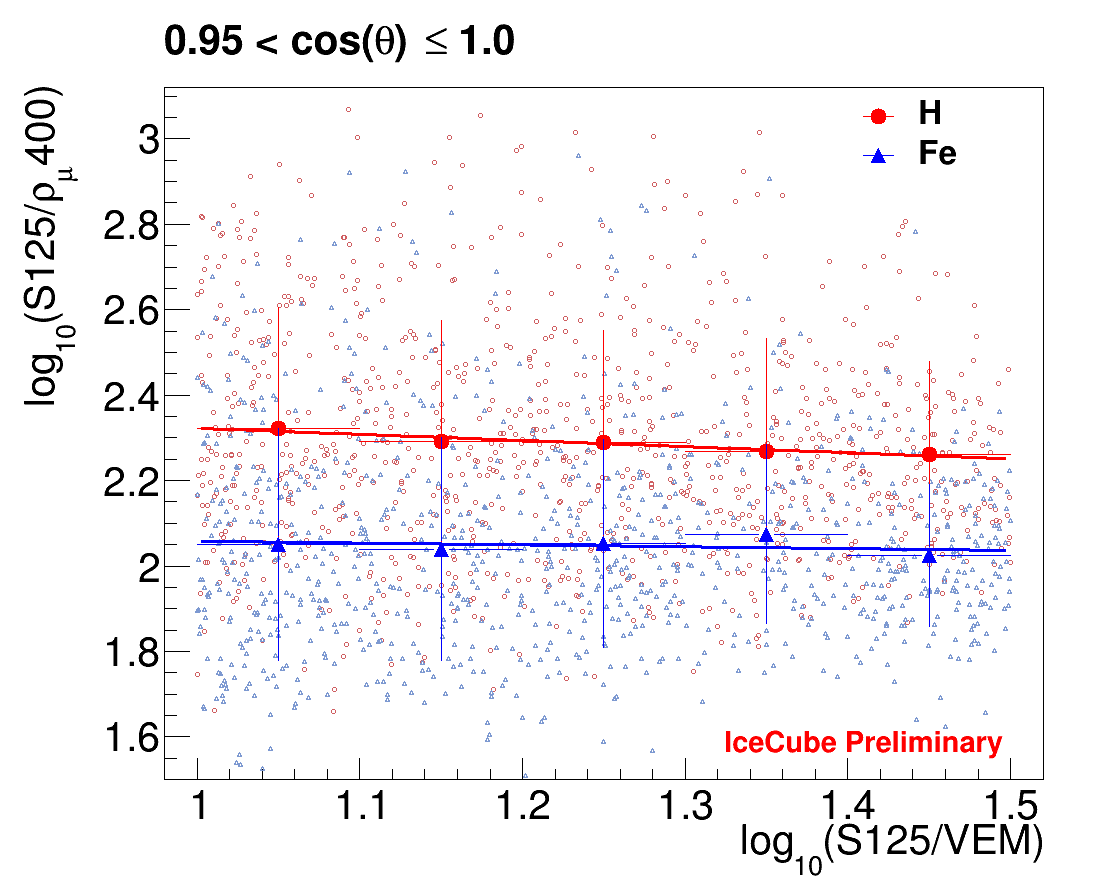}
    \includegraphics[width=0.49\textwidth]{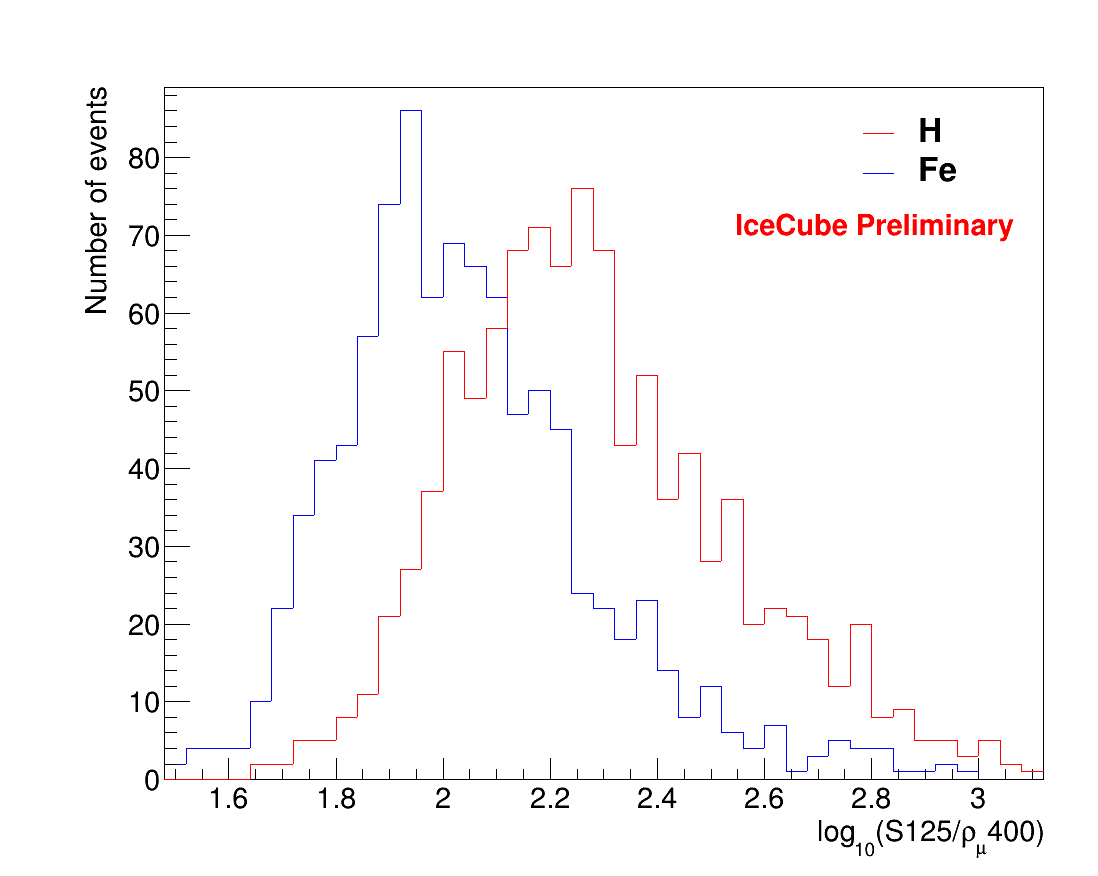}
    \caption{Left: Ratio between the shower size S125 and the muon-based parameter $\rho_{\mu}400$ as a function of S125 for proton and iron assumptions. The lines of best fit are shown as well.
    Right: Its projection onto the y-axis.}
\label{fig6}
\end{center}
\end{figure}
Figure 6 shows the fraction of the estimated muon-based parameter $\rho_{\mu}400$ with the muon correction to the shower size S125 for proton and iron assumptions (left), shown with the projection onto the y-axis (right).

\section{Result}
By means of the correlation of the shower size and the estimated muon parameter, the muon-based mass sensitive parameter is estimated. It is often called the normalized shower size, $k$ parameter \cite{KG}, which is defined as:
\begin{equation}
  k = \frac{\rm{log}_{10}(\frac{S125}{\rho_{\mu,400,data}}) - \rm{log}_{10}(\frac{S125}{\rho_{\mu,400}})_{H}}
  {\rm{log}_{10}(\frac{S125}{\rho_{\mu,400}})_{Fe} - \rm{log}_{10}(\frac{S125}{\rho_{\mu,400}})_{H}},
\end{equation}

\begin{equation}
  \rm{log}_{10}(\frac{S125}{\rho_{\mu,400}})_{H,Fe} = c_{H,Fe} \cdot \rm{log}_{10}(S125) + d_{H,Fe}.
\end{equation}
The coefficients $c_{H,Fe}, d_{H,Fe}$ were determined by fitting the distribution of the normalized shower size ratio $S125/\rho_{\mu}400$ (Fig. 6).
The coefficients are $c_{H} = -0.15 \pm 0.06, d_{H} = 2.47 \pm 0.07$ and $c_{Fe} = -0.05 \pm 0.04, d_{Fe} = 2.11 \pm 0.06$ for proton and iron primaries, respectively. 

\begin{figure}[t!]
  \begin{center}
    \includegraphics[width=0.47\textwidth]{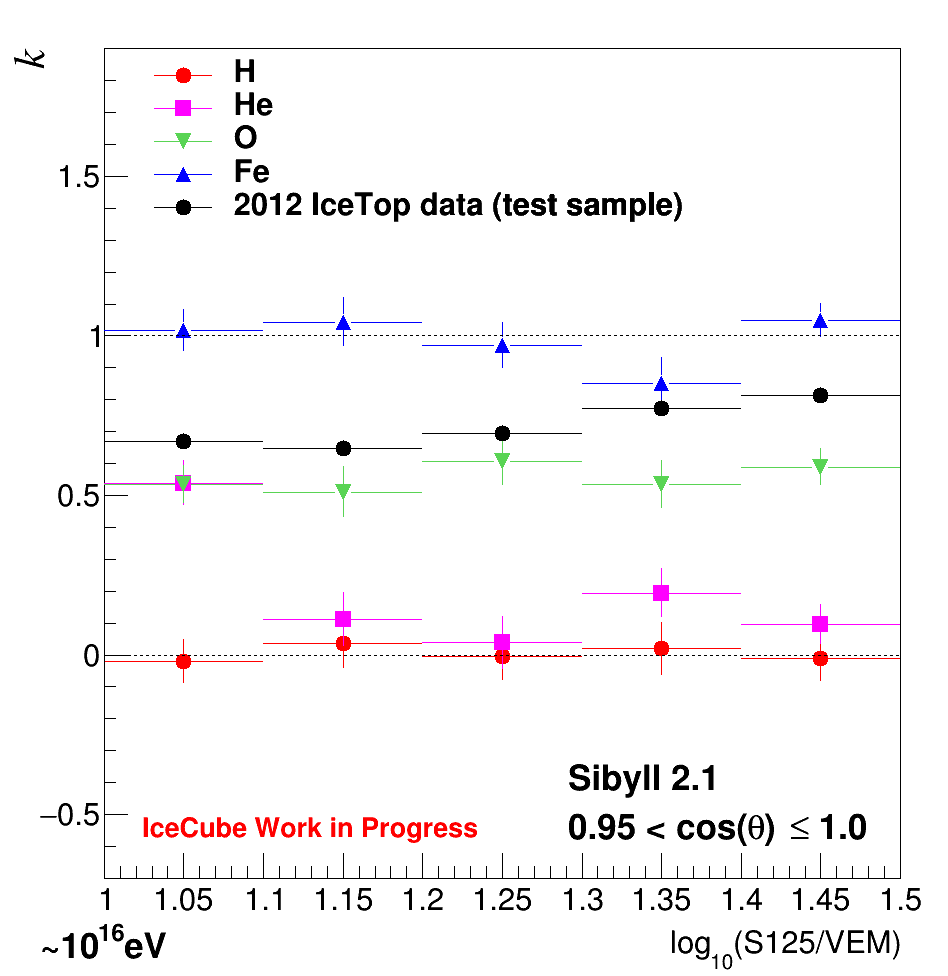}
    \caption{Muon-based mass sensitive parameter $k$ as a function of the energy estimator (S125) for experimental data compared with simulations of four different primary masses based on the hadronic interaction model of Sibyll 2.1. Only statistical error bars are shown.}
\label{fig7}
\end{center}
\end{figure}
Figure 7 shows the evolution of the $k$ parameter as a function of S125 in the primary energy range of $10^{16}$ to $10^{16.5}$ eV. This distribution reveals clearly the mass dependence. From the definition of the $k$ parameter, the value of $k$ is 0 for the assumption of proton showers and 1 for iron showers. The intermediate primary particles are between 0 and 1. The data seems to favor the heavy primaries, as we expected in this energy range. This parameter can be used to separate the events into the different mass groups.
The first energy bin for showers induced by primary helium nuclei has a higher $k$ value than the other bins. It might be related with a bias of energy estimator, however, a detailed investigation is ongoing.

\section{Conclusion and outlook}
Considering the charge signal distributions measured by IceTop, a muon enriched parameter ($\rho_{\mu}400$) per individual shower was defined and estimated. Using this parameter and the shower size S125, the energy-dependent mass sensitive ratio $k$ was estimated event by event.
The study also showed that this analysis method seems to work properly for the IceTop data.
This mass sensitive parameter can be applied for mass composition and, in combination with S125, for reconstruction of energy spectrum of primary cosmic rays \cite{KG}. This work is in progress.

\clearpage
\section*{Full Author List: IceCube Collaboration}

\scriptsize
\noindent
R. Abbasi$^{17}$,
M. Ackermann$^{59}$,
J. Adams$^{18}$,
J. A. Aguilar$^{12}$,
M. Ahlers$^{22}$,
M. Ahrens$^{50}$,
C. Alispach$^{28}$,
A. A. Alves Jr.$^{31}$,
N. M. Amin$^{42}$,
R. An$^{14}$
% ...
\\

\noindent
% ...
$^{12}$ Universit{\'e} Libre de Bruxelles, Science Faculty CP230, B-1050 Brussels, Belgium \\
% ...
$^{14}$ Department of Physics and Laboratory for Particle Physics and Cosmology, Harvard University, Cambridge, MA 02138, USA \\
% ...
$^{17}$ Department of Physics, Loyola University Chicago, Chicago, IL 60660, USA \\
$^{18}$ Dept. of Physics and Astronomy, University of Canterbury, Private Bag 4800, Christchurch, New Zealand \\
% ...
$^{22}$ Niels Bohr Institute, University of Copenhagen, DK-2100 Copenhagen, Denmark \\
% ...
$^{28}$ D{\'e}partement de physique nucl{\'e}aire et corpusculaire, Universit{\'e} de Gen{\`e}ve, CH-1211 Gen{\`e}ve, Switzerland \\
% ...
$^{31}$ Karlsruhe Institute of Technology, Institute for Astroparticle Physics, D-76021 Karlsruhe, Germany  \\
% ...
$^{42}$ Bartol Research Institute and Dept. of Physics and Astronomy, University of Delaware, Newark, DE 19716, USA \\
% ...
$^{50}$ Oskar Klein Centre and Dept. of Physics, Stockholm University, SE-10691 Stockholm, Sweden \\
% ...
$^{59}$ DESY, D-15738 Zeuthen, Germany \\
% ...

\subsection*{Acknowledgements}

\noindent
USA {\textendash} U.S. National Science Foundation-Office of Polar Programs,
U.S. National Science Foundation-Physics Division,
U.S. National Science Foundation-EPSCoR,
Wisconsin Alumni Research Foundation,
Center for High Throughput Computing (CHTC) at the University of Wisconsin{\textendash}Madison,
Open Science Grid (OSG),
Extreme Science and Engineering Discovery Environment (XSEDE),
Frontera computing project at the Texas Advanced Computing Center,
U.S. Department of Energy-National Energy Research Scientific Computing Center,
Particle astrophysics research computing center at the University of Maryland,
Institute for Cyber-Enabled Research at Michigan State University,
and Astroparticle physics computational facility at Marquette University;
Belgium {\textendash} Funds for Scientific Research (FRS-FNRS and FWO),
FWO Odysseus and Big Science programmes,
and Belgian Federal Science Policy Office (Belspo);
Germany {\textendash} Bundesministerium f{\"u}r Bildung und Forschung (BMBF),
Deutsche Forschungsgemeinschaft (DFG),
Helmholtz Alliance for Astroparticle Physics (HAP),
Initiative and Networking Fund of the Helmholtz Association,
Deutsches Elektronen Synchrotron (DESY),
and High Performance Computing cluster of the RWTH Aachen;
Sweden {\textendash} Swedish Research Council,
Swedish Polar Research Secretariat,
Swedish National Infrastructure for Computing (SNIC),
and Knut and Alice Wallenberg Foundation;
Australia {\textendash} Australian Research Council;
Canada {\textendash} Natural Sciences and Engineering Research Council of Canada,
Calcul Qu{\'e}bec, Compute Ontario, Canada Foundation for Innovation, WestGrid, and Compute Canada;
Denmark {\textendash} Villum Fonden and Carlsberg Foundation;
New Zealand {\textendash} Marsden Fund;
Japan {\textendash} Japan Society for Promotion of Science (JSPS)
and Institute for Global Prominent Research (IGPR) of Chiba University;
Korea {\textendash} National Research Foundation of Korea (NRF);
Switzerland {\textendash} Swiss National Science Foundation (SNSF);
United Kingdom {\textendash} Department of Physics, University of Oxford.

\end{document}